# Robust Cooperative Spectrum Sensing for Disaster Relief Networks in Correlated Environments

Nuno Pratas[1,2], Nicola Marchetti[1], Neeli Rashmi Prasad[1], António Rodrigues[2] and Ramjee Prasad[1]

*Abstract*—Disaster relief networks are designed to be adaptable and resilient so to encompass the demands of the emergency service. Cognitive Radio enhanced ad-hoc architecture has been put forward as a candidate to enable such networks.

Spectrum sensing, the cornerstone of the Cognitive Radio paradigm, has been the focus of intensive research, from which the main conclusion was that its performance can be greatly enhanced through the use of cooperative sensing schemes. To apply the Cognitive Radio paradigm to Ad-hoc disaster relief networks, the design of effective cooperative spectrum sensing schemes is essential.

In this paper we propose a cluster based orchestration cooperative sensing scheme, which adapts to the cluster nodes surrounding radio environment state as well as to the degree of correlation observed between those nodes. The proposed scheme is given both in a centralized as well as in a decentralized approach. In the centralized approach, the cluster head controls and adapts the distribution of the cluster sensing nodes according to the monitored spectrum state. While in the decentralized approach, each of the cluster nodes decides which spectrum it should monitor, according to the past local sensing decisions of the cluster nodes. The centralized and decentralized schemes can be combined to achieve a more robust cooperative spectrum sensing scheme.

The proposed scheme performance is evaluated through a framework, which allows measuring the accuracy of the spectrum sensing cooperative scheme by measuring the error in the estimation of the monitored spectrum state.

Through this evaluation it is shown that the proposed scheme outperforms the case where the choice of which spectrum to sense is done without using the knowledge obtained in previous sensing iterations, i.e. a implementation of a blind Round Robin scheme.

## I. INTRODUCTION

A disaster, according to [1], can be loosely defined as an event, which over a relatively short time period, causes a large number of casualties and infrastructure damages. In a scenario where the breakdown of the communication infrastructure occurs, it is vital that the communications between the relief groups, upon network deployment, are established as quickly and as easily as possible. This raises the need for specifically tailored infrastructure-less mobile networks with the ability to dynamically access the available radio spectrum. To accomplish this, it can be considered to combine in one network the Mobile Ad-hoc architecture and Cognitive Radio (CR) paradigms [2,3]. The cornerstone of such network is its ability to adapt to the surrounding environment, which can only be accomplished if accurate information of it is made available.

In this paper we focus on the spectrum sensing part of environment awareness. Spectrum sensing, covered extensively in literature [4-5], falls into what is known as detection theory, presented in detail in [6]. Spectrum sensing is realized as a physical and MAC layer mechanism. The physical layer sensing focuses on detecting signals, and the detection methods put in place can be classified into two groups, either coherent (prior information needed, e.g. Pilot Detection, [4]) or non-coherent (no prior information needed, e.g. Energy Detector, [7]). The MAC layer part of the spectrum sensing - the target of this paper - focuses on when to sense and which spectrum to sense.

Considering that the sensing requirements are set by the channel conditions, which depend on the path loss, multipath, shadowing and local interference, the combination of these phenomena can result in regimes where the signal SNR is below the detection threshold of the sensor. To overcome this limitation, in [8–11] it was proposed the use of cooperation in the spectrum sensing. Since the signal strength varies with the sensor location, the worst fading conditions can be avoided if multiple sensors in different spatial locations share their sensing measurements, i.e. take advantage of the spatial diversity. Most of these proposed cooperative methods are based on data fusion techniques to perform the decision on what is the actual state of the spectrum. Recently, in [12], it has been proposed the use of a scheduling scheme to select which channels to sense based on the channel statistics, although the authors assumed that they had prior information on the correct channel occupation statistics. On [13] the authors did a comprehensive study on the effect of correlation on cooperative spectrum sensing, besides providing a theoretical framework based on Bayesian inference, the paper's main conclusion was that under certain correlation conditions the use of cooperation is not worthwhile.

In this paper a cooperative spectrum sensing scheme is proposed for an Ad-hoc based scenario. The proposed scheme works under the assumption that all the nodes in this ad-hoc network are grouped in clusters. The proposed scheme allows assigning a channel to sense to each of the cluster nodes. This assignment is done according to previously estimated channel occupation statistics. The proposed scheme is accomplished following both a centralized and a decentralized approach. In the centralized approach, the cluster head is responsible for collecting and fusing the local decisions of each of the cluster nodes, with the purpose of estimating the monitored channels occupancy. The estimated channel occupations statistics are then used by the cluster head to assign to each cluster node which channel to sense. In the cluster decentralized approach, each one of the cluster nodes gathers and fuses the local decisions of the other cluster nodes and from there it estimates the monitored channels occupancy. Using the estimated information the cluster node decides which channel to sense in the next sensing session.

The remainder of this paper is organized as follows. In Section II it is given the problem definition together with a short description of the methodology used to solve it, and in Section III the system design is discussed. In Section IV one studies the data fusion scheme under certain correlation conditions for the cooperative spectrum sensing, while in Section V two channel state estimators are considered. In Section VI one proposes and discusses in details the sensing orchestration schemes, whereas, in Section VII a comparative performance evaluation of the integrated proposed schemes is presented. Finally, Section VIII concludes the paper with a recap of the contribution and of the main obtained results, as well as an outlook on further development.

## II. PROBLEM DEFINITION AND METHODOLOGY

### A. Problem Definition

The problem we tackle in this paper is the following: *Consider an Ad-hoc Cognitive Radio disaster relief network, composed of network nodes, organized in clusters, and capable of operating on and sensing any narrow band channel of a targeted range of spectrum. How should these cluster nodes cooperate, so that at any given time all the cluster nodes have accurate statistics of the targeted spectrum?*

To tackle this problem we present a MAC layer distributed spectrum sensing mechanism. The mechanism was developed following a centralized as well as a decentralized approach, both elaborated in the later sections of the paper.

The proposed distributed mechanism, in both approaches, allows for each of the sensing nodes to sense any channel of the monitored spectrum while ensuring that at cluster level a detection, as accurate as possible, of the available spectrum is achieved in all nodes.

### B. Methodology

The methodology followed mirrors the article structure, and is the following: first we describe the system design and scenario assumptions, then we identify from the system design the three topics that are treated in this paper, being these: Spectrum Sensing Results Fusion, Channel State Estimation and Sensing Orchestration. Each of these topics is analyzed in depth in a dedicated section.

The study then concludes with the integration of the analyzed topics in a distributed spectrum sensing scheme and a performance comparison between the centralized and decentralized approaches.

### C. Evaluation Metrics

The goal of spectrum sensing, cooperative or not, is to find out what is the state of the channel being sensed. This process, illustrated in Figure 1, can be described as a mechanism which corresponds to an imperfect and simplified mapping of the real radio environment to a representation in the sensing node. Therefore a metric or a set of metrics that can gauge how imperfect/accurate a spectrum sensing method is needed.

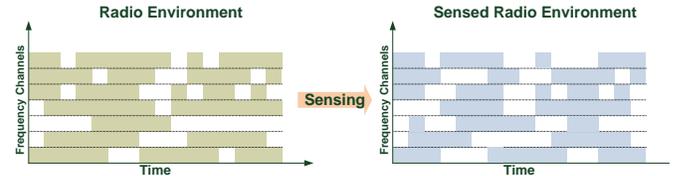

**Figure 1 - Spectrum sensing as a mapping mechanism.**

The Root Mean Square Error (RMSE) is frequently used to measure the differences between values predicted by a model or an estimator and the values actually observed from the phenomenon being modelled or estimated; therefore RMSE is a measure of accuracy. While considering that $\hat{s}_{m,i}$ represents the channel's *m* estimated mean un-occupancy in sensing session *i* and $s_m$ is the channel's *m* real mean un-occupancy. Then the individual accounted differences, $\hat{s}_{m,i} - s_m$, are called residuals, and the RMSE aggregates these individual residuals into a single measure of prediction.

Here we extend the RMSE to measure the accuracy of the proposed sensing scheme, by applying it to the estimated mean channel un-occupancy. The RMSE of the channel *m* estimated un-occupancy is given by,

$$RMSE(\hat{s}_m) = \sqrt{\frac{\sum_{i=1}^{N}(\hat{s}_{m,i}-s_m)^2}{N}} \quad (1)$$

where *N* is the number of sensing sessions.

This metric is used to gauge the performance of the *n* channels with the greater un-occupancy, being incorporated in the Multi-Channel Effective Root Mean Square Error, $RMSE_{ME}$, which is defined as,

$$RMSE_{ME}(n) = \frac{\sum_{i \in \Omega} RMSE(\hat{s}_m)}{n} \quad (2)$$

where $\Omega$ is the set of *n* channels with greater un-occupancy. The $RMSE_{ME}(n)$ metric allows to gauge how accurate is the estimation of the channel mean un-occupancy in the *n* top un-occupied channels.

## III. System Design

### A. Preliminaries

It is assumed that the targeted spectrum for sensing is divided in channels of equal bandwidth, and that each of the network nodes is able to sense only one of these channels. Furthermore, it is assumed that the network nodes are already organized in clusters.

Each node of the network is assumed to have available two logical types of channel, the Control Channel (CCH) where all control information is exchanged and the Data Channel (DCH) through which all users data are exchanged. For the physical implementation of these channels it is assumed that they have independent transceivers associated.

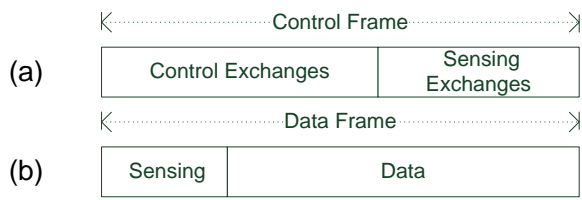

Figure 2 - (a) CCH Control Frame, (b) DCH Data Frame.

The CCH method of access is CSMA/CA. The spectrum channel allocated to the CCH is assumed to be cluster-dedicated, i.e., the channel contenders are the other cluster nodes; the CCH associated transmission time is divided as depicted in Figure 2-(a). The control exchanges are fully dedicated to control functions (transmit and receive control data), while the sensing exchanges are dedicated to the report of the sensing as well as of the orchestration decisions (in the case of the centralized scheme).

The DCH access method is out of scope. The only assumption is that part of the reception/transmission time is dedicated periodically to spectrum sensing, as depicted in Figure 2-(b). The DCH is able to operate and sense dynamically in different spectrum channels, one for each sensing session, whereas these channels are assumed to have the same bandwidth.

We consider that each cluster node performs the sensing through the use of a Energy Detector (ED) [7-8], an non-coherent sensing scheme. The ED was chosen since the conditions of the radio spectrum environment and of its incumbent signals structure are unknown to our network, therefore only non-coherent schemes are considered, since due to lack of prior information coherent spectrum sensing schemes are rendered useless.

### B. Common System Design

The purpose of the distributed spectrum sensing mechanism is to ensure that all of the cluster nodes have updated and synchronized information about the state of the targeted spectrum.

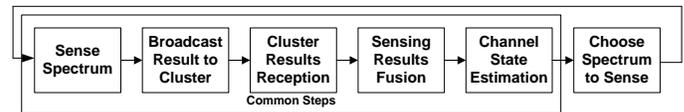

Figure 3 - Distributed spectrum sensing mechanism flow.

The distributed spectrum sensing mechanism flow is depicted in Figure 3 in different steps:

- **Spectrum Sensing** - Each cluster node performs the sensing through the use of a Energy Detector, therefore at the end of the spectrum sensing a binary decision regarding the status of the sensed channel is taken by each node. The sensing occurs in the DCH in the "Sensing" period depicted in Figure 2-(b);
- **Broadcast Result to Cluster** - Each cluster node shares the result of the binary decision reached in the spectrum sensing. The sharing is done through broadcast, which is done through the CCH during the *Sensing Exchanges* period depicted in Figure 2-(a);
- **Cluster Results Reception** - Each cluster node receives the results broadcasted by the remaining cluster nodes through the CCH;
- **Sensing Results Fusion** - At this point, the node fuses together the sensing results received from the other cluster nodes. Note that the fusion process is done separately for each sensed channel. The data fusion scheme used is classified as a synchronous hard decision, and therefore the result is a binary decision;
- **Channel State Estimation** - The estimation of the channels state is done based on past observations and current observations, when available, i.e. if the channel in question was sensed. Through this process it is possible to obtain updated statistics of the network targeted channels;
- **Choose Spectrum to Sense** - This step depends on the approach chosen to implement the mechanism, i.e. if the mechanism is centralized or decentralized coordination. In the centralized approach one of the cluster nodes decides which channel should be sensed by each of the cluster nodes, while in the decentralized approach, the decision on which channel to sense is done independently by each of the cluster nodes. Both approaches perform this choice according to the channels occupation statistics.

The flows of the centralized and decentralized mechanism implementation have in common all the steps except for the "Choose Spectrum to Sense" step. The differences of each implementation will be depicted in the following sub-sections.

### C. Centralized System Design

The centralized distributed spectrum sensing mechanism steps are depicted in Figure 4, and in Figure 5 are depicted the "Sensing Exchanges" in the CCH.

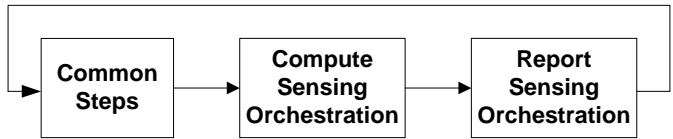

Figure 4 - Centralized Mechanism.

The Common Steps were the ones described in the previous section, the remaining ones are:
- **Compute Sensing Orchestration** - In this step the coordinating node, i.e. the cluster head, computes what will be the distribution of the sensing nodes in the following sensing sessions, a process which we call orchestration. This process is implemented through a resource scheduler, which will be presented in a later section of the paper. This step occurs during the "Processing" period depicted in Figure 5.
- **Report Sensing Orchestration** - After the orchestration has been computed, the cluster head reports it to the cluster nodes. This report is broadcasted through the CCH in the "Orchestration Reporting" period depicted in Figure 5.

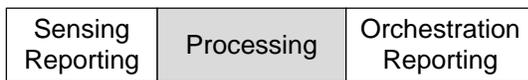

Figure 5 - Centralized Mechanism "Sensing Exchanges" period of Figure 2-(a),.

### D. Decentralized System Design

The decentralized distributed spectrum sensing mechanism steps are depicted in Figure 6. Since in the decentralized case the next channel to be sensed is chosen by the node itself, then the "Sensing Exchanges" period of the CCH depicted in Figure 2-(a) is used only by the cluster nodes to report their sensing results.

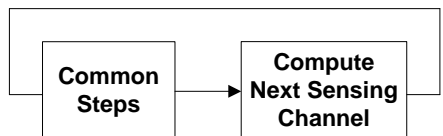

Figure 6 - Decentralized Mechanism.

The choice of which channel the cluster node should sense in the next sensing session is done following the same approach as the centralized mechanism, i.e. it is based on a resource scheduler, which will be presented in a later section of the paper.

### E. Considerations

The distributed spectrum mechanism steps were described for both the centralized and the decentralized approach. The first remark is that the centralized scheme requires more radio resources due to the orchestration reporting. But the tradeoff of the decentralized scheme is that the resource distribution, i.e. the nodes sensing the channels in next sensing session, might not be as optimal as in the centralized scheme.

So the choice between these schemes is between optimality of resource allocation, and consequentially energy and radio resources spent due to extra signaling, or sub-optimal resource allocation.

From the system description were identified three tasks which will be focused in this paper, being those:
- **Spectrum Sensing Results Fusion** - The data fusion scheme considered is the synchronous hard decision, which uses predefined fusion rules to achieve the data fusion. In this paper we will derive a data fusion rule which will take into account the properties of the phenomenon being observed, i.e. of the radio channel, and the correlation observed between the cluster nodes;
- **Channel State Estimation** - The estimation of channel state is done based on past observations when available. In this paper we will derive an estimator which accomplishes this;
- **Sensing Orchestration** - The sensing orchestration scheme which distributes the cluster nodes across the targeted spectrum is derived. This will be extended to accomplish the orchestration in the centralized approach as well as in decentralized approach.

These three tasks will be discussed in the next sections.

## IV. SPECTRUM SENSING RESULTS FUSION

### A. Introduction to data fusion

Cooperative sensing is a distributed detection system specialized in detecting the state of the monitored spectrum. In this paper we consider a two-level distributed detection system, as depicted in Figure 7. This system consists of a number of local detectors and a fusion center. The local detectors make a decision of the underlying binary hypothesis testing problem based on their own observations and then transmit their decisions to the fusion center where the global decision is derived. In this paper we consider that the local detectors were already designed to achieve a certain detection performance, i.e. we assume that they are Energy Detectors designed to detect signals above a certain SNR. We also consider that these detectors have an homogeneous performance, i.e. all the detectors were designed to detect the same level of SNR.

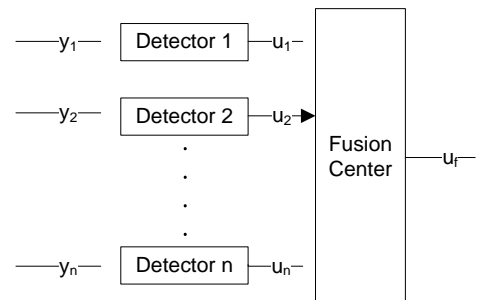

Figure 7 - Two-level distributed detection system.

In the fusion center the local decisions need to be combined so that a global decision is achieved. In this section we describe the design of such fusion rule. In the fusion center a binary hypothesis testing is done, and it is through this testing that the fusion rule is defined. In the literature two approaches to model this problem are commonly followed, [6], the Bayesian approach (were the conditional densities under the two hypothesis have to be priory known as well as the cost of the action that can be followed, which in a two hypotheses problem is four), and the Neyman-Pearson (N-P) test where such information is not needed. Here we follow the N-P test formulation, since we assume that we do not have access to any prior knowledge about the signals that we are trying to detect.

*B. Data fusion formulation*

Consider the binary hypothesis testing problem with hypothesis $H_0$ and $H_1$, which correspond respectively to the cases where the sensed channel is vacant and occupied. The fusion center implements the N-P test by using all the decisions that the individual sensors have communicated. The N-P test, $\Lambda(\boldsymbol{u})$, is formulated as the following Likelihood Ratio (LR) test:

$$\Lambda(\boldsymbol{u}) = \frac{P(u_1, u_2, \cdots, u_n | H_1)}{P(u_1, u_2, \cdots, u_n | H_0)} \underset{H_0}{\overset{H_1}{\gtrless}} \lambda \qquad (3)$$

where $\boldsymbol{u} = [u_1 ... u_n]^T$, is the vector formed by the set of local decisions ($u_i$ takes the value 0 when the sensor $i$ decides $H_0$, and $u_i = 1$ when the sensor $i$ decides $H_1$) corresponding to the $n$ local detectors, and $\lambda$ is the threshold determined by setting an upper bound to the probability of false alarm at the fusion center, [6].

Considering the case were all the local detectors/sensors have the same performance and that there is a quantifiable degree of correlation between their decisions, then according to [14], the Likelihood Ratio Test (LRT) to that case can be defined as,

$$\Lambda(u) = \frac{P(u|H_1)}{P(u|H_0)} = \Lambda(m) \qquad (4)$$

where $m$ out of $n$ detectors are in favor of $H_0$ (i.e. there are $m$ 0's in the vector $\boldsymbol{u}$). Elaborating further $\Lambda(\boldsymbol{u})$, [14] we get,

$$\Lambda(\boldsymbol{u}) = \frac{\sum_{i=0}^{m}(-1)^i \binom{m}{i} P_d \prod_{k=0}^{n-m+i-2} \frac{\rho_1(k+1-P_d) + P_d}{1+k\rho_1}}{\sum_{i=0}^{m}(-1)^i \binom{m}{i} P_{fa} \prod_{k=0}^{n-m+i-2} \frac{\rho_0(k+1-P_{fa}) + P_{fa}}{1+k\rho_0}},$$

$$0 \leq m \leq n-2 \qquad (5)$$

where $P_d$ and $P_{fa}$, are respectively the probability of detection and probability of false alarm, and the correlation indices $\rho_0$ and $\rho_1$, where $0 \leq \rho_0 \leq 1$ and $0 \leq \rho_1 \leq 1$, are defined by,

$$\rho_k = \frac{E[u_i u_j | H_k] - E[u_i | H_k] E[u_j | H_k]}{\sqrt{E[(u_i - E[u_i])^2 | H_k] E[(u_j - E[u_j])^2 | H_k]}}, \forall_{i,j}\ i \neq j, k=0,1 \qquad (6)$$

$\Lambda(\boldsymbol{u})$ for the case when $m = n-1$ and $m = n$ was not explicitly considered in [14], but were derived in [15].

The optimum rule, i.e. the one maximizing the global probability of detection, for a given upper bound of the global probability of false alarm is obtained by the LRT, given by,

$$rule_{optimal}(\boldsymbol{u}) = rule_{optimal}(m) =$$
$$\begin{cases} H_1 & , if\ \Lambda(m) > \lambda \\ H_1 with\ probability\ \gamma & , if\ \Lambda(m) = \lambda \\ H_0 & , if\ \Lambda(m) < \lambda \end{cases} \qquad (7)$$

where $\gamma$ is a randomization constant, and $\lambda > 0$ and $\gamma > 0$ are defined according to the global probability of false alarm upper bound.

Considering that the implementation of the LRT leads to complex iterative algorithms, and that in [14] it was shown that the LRT can be expressed as a function of $m$, i.e. the number of detectors that decide in favor of $H_0$, in [15] was proposed the use of a counting rule, i.e. a rule that counts $m$, and decide $H_1$ when $m$ is smaller than a given integer threshold. Therefore a counting rule can be defined by,

$$rule_{count}(\boldsymbol{u}) = rule_{count}(m) =$$
$$\begin{cases} H_1, & if\ m < m_0 \\ H_1\ with\ probability\ \gamma, & if\ m = m_0 \\ H_0, & if\ m > m_0 \end{cases} \qquad (8)$$

The equivalence between the LRT and the counting rule, i.e., $\Lambda(m) \lesseqgtr \lambda \Leftrightarrow m \lesseqgtr m_0(\lambda)$ is only valid if $\Lambda(m)$ is a decreasing function of $m$, which was thoroughly demonstrated in [15], i.e. it was shown that for properly operating local detectors ($P_d >> P_{fa}$) the counting rule is almost a Uniformly Most Powerful (UMP) test, [15].

The choice of the counting rule threshold, $m_0$, depends on the local detectors performance, the correlation between the decisions of these detectors and finally on the upper bound set for the global probability of false alarm. But considering that the correlation between the decisions change over time, when the detectors are mobile, then there should a mechanism which adapts the $m_0$ to these changing conditions. We propose such a mechanism in the next sub-section.

*C. Data fusion through adaptive counting rule mechanism*

The need for an adaptive counting rule can be motivated by the example given in Figure 8 and Figure 9, where it is depicted the RMSE of the decision at the fusion center compared to the real state of the signal, according to $k$, the decision threshold and $k = n - m_0$. From the figures we observe that the optimum $k$ depends on the average correlation index between the local detectors. Therefore there must be a mechanism which allows selecting the optimum $k$ according to the correlation index. Note that the optimum $k$ depends also on the sensors performance as well as the number of sensors put in place to sense the channel.

Here we propose a mechanism which adapts continuously the $k$ according to the feedback from the previous global decisions, i.e. if the sensed channel is deemed available then the cluster nodes try to use the channel, and the result of that action is used as feedback to tune the *k*. If the channel is occupied then it means that there was a misdetection, and therefore it means the *k* should be decreased, increasing the probability of detection. At the same time there should be an opposite mechanism which increases *k*, so to ensure that the upper bound of the global false alarm probability is respected.

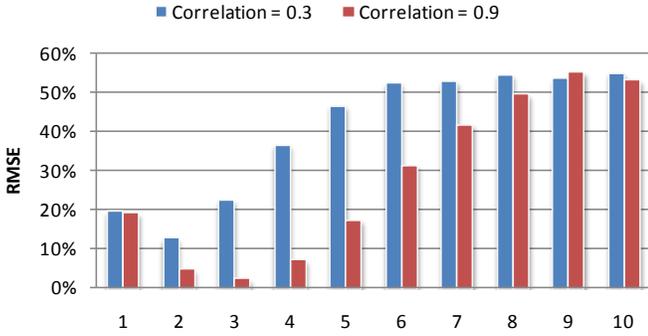

Figure 8- RMSE vs k according to correlation index, with 10 local detectors.

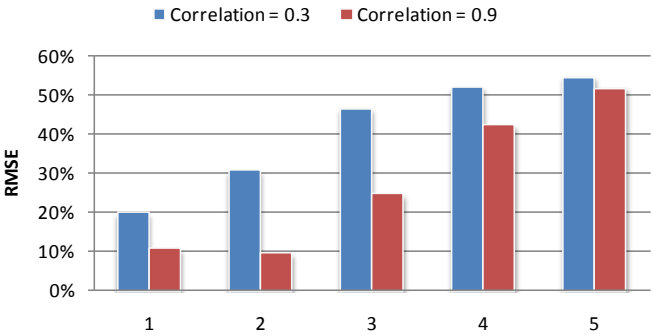

Figure 9 - RMSE vs k according to correlation index, with 5 local detectors.

The proposed algorithm makes use of the following quantities:
- $MD_C$ - The Misdetection Counter counts every time the cluster decides that the channel is free, and the network tries to use the channel but the channel is occupied;
- $OC_C$ - The Occupied Detection Counter counts every time the cluster decides the channel is occupied;
- $O_W$ - The Observation Window is the time interval of the observations used to make the decisions, i.e. it is the memory of what occurred in the previous sensing sessions;
- $MD_{Thrs}$ - The Misdetection Threshold refers to the amount of misdetection permitted during the observation window. When this value is overcome triggers the *k* needs to be decreased. This threshold is obtained by multiplying the upper bound of the global probability of false alarm with $O_W$;
- $OC_{Thrs}$ - The Occupied Detection Threshold refers to the amount of occupied detections occurring during the $O_W$, which if overcome triggers the increase of the needed positive detections to decide a global positive detection. It is obtained by multiplying the $O_W$ with the upper bound of the allowed global false alarm probability.

The algorithm which implements the adaptable counting rule is depicted in Algorithm 1. This algorithm adapts the decision threshold *k*, so that for a detection to occur there have to be at least *k* local detections out of the *n* local detector.

```
%After the Fusion Centre receiving the local decisions
Compute current decision, U_F
If (U_F is H_1)
    Increment OC_C
Else
    If (Channel state is not H_0)
        Increment MD_C
If (MD_C>MD_Thrs)
    Decrease k
If (OC_C > OC_Thrs)
    Increase k
Update MD_C and OC_C %discard observations outside the O_W
%End of Adaptive Counting Rule
```

**Algorithm 1 - Adaptive counting rule mechanism.**

*D. Adaptive counting rule performance comparison*

To illustrate the performance of the adaptive counting rule mechanism, we compare it with two extreme cases, the OR rule (*1*-out-of-*n*) and the AND rule (*n*-out-of-*n*), while considering the same scenario it was used to obtain the results shown in Figure 8 and Figure 9.

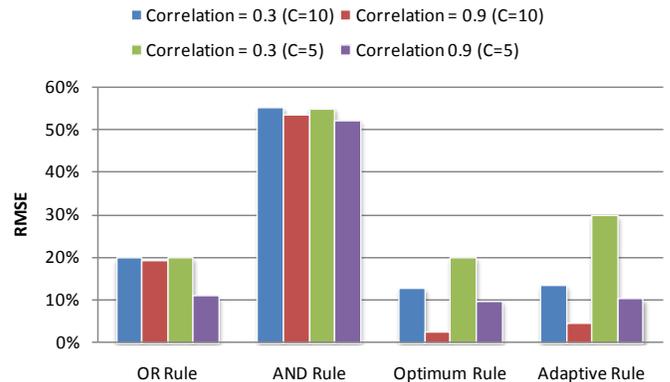

Figure 10 - RMSE according to fusion rule.

The RMSE obtained while using the different rules is depicted in Figure 10. As expected the adaptive rule RMSE is close to the one obtained with the optimum rule, while outperforming both the OR and AND rule, except for the case where correlation is 0.3 with C=5. In this case the performance of the adaptive rule was worse than the optimum/OR rule because it is unable to maintain the *k*, due to its dynamic nature, equal to the value of the lower bound (*k=1*) or upper bound (*k=n*).

The proposed adaptive fusion rule algorithm allows achieving almost the same performance as the optimum rule, without needing as input the correlation between the local detectors.

## V. CHANNEL STATE ESTIMATION

### A. Introduction

After the sensing of the channel is done and its current state determined, it is time to combine the current observed state with past observations: through this process it is then possible to obtain updated statistics of the sensed channel state, e.g. average occupation, free period, etc. The use of these statistics can then allow for a more adequate decision on when to use the channel for data transmission.

Here we consider the estimation of the channel's $m$ mean un-occupancy, denoted as $s_m$, which is estimated continuously during the network lifetime. This estimation is based on previous observations, when available, while taking in account when a channel has not been observed in a long time. The $s_m$ is the estimation of $P(H_0)$.

To estimate $s_m$ we study two methods, which can be classified as an exponential moving average and a linear moving average, respectively. Both include a reset mechanism which allows updating the $s_m$, although introducing an error, even when the channel was not sensed. The need for this mechanism, is that since the $s_m$ is used as a bid in the orchestration scheme, then there needs to be a feedback mechanism which enables a controlled distribution of the sensing nodes by using the $\hat{s}_m$ as the channel bid, i.e. $w_m$. This will be made clearer in Section VI.

### B. Exponential Moving Average Estimator

The exponential moving average with reset factor channel un-occupancy estimator is given by,

$$\hat{s}_m = \begin{cases} (1-\alpha)\hat{s}_{m,old} + \alpha\, s_{m,inst}, & \text{if } m \text{ sensed} \\ (1-\alpha)\hat{s}_{m,old} + \alpha\, s_{reset}, & \text{if } m \text{ not sensed} \end{cases} \quad (9)$$

where $\hat{s}_m$ is the estimated un-occupancy of the channel $m$, $\hat{s}_{m,old}$ is the estimated un-occupancy from the last spectrum sensing session, $s_{m,inst}$ is the instantaneous occupancy obtained from the sensing of the channel $m$ in the previous sensing session, $s_{reset}$ is the term that is used to reset the channel un-occupancy estimation when the channel has not been sensed in the previous sensing session and finally $\alpha$ is the forgetting factor used to tune the exponential moving average smoothness.

### C. Linear Moving Average Estimator

The linear moving average with reset factor channel un-occupancy estimator is given by,

$$\hat{s}_m = \begin{cases} \dfrac{\sum_{i=1}^{O_W} s_{m,old}^i + s_{m,inst}}{O_W}, & \text{if } m \text{ sensed} \\ \dfrac{\sum_{i=1}^{O_W} s_{m,old}^i + s_{reset}}{O_W}, & \text{if } m \text{ not sensed} \end{cases} \quad (10)$$

where $\hat{s}_m$ is the estimated un-occupancy of the channel m is, $s_{m,old}^i$ is the observed un-occupancy in the $i^{th}$ previous spectrum sensing session, $s_{m,inst}$ is the instantaneous occupancy obtained from the sensing of the channel m in the previous sensing session, $s_{reset}$ is the term that is used to reset the channel un-occupancy estimation when the channel has not been sensed in the previous sensing session, and $O_W$ is the observation window length.

### D. Implementation Issues and Performance Comparison

The linear estimator achieves the minimum RMSE if the observation window is the same as the number of observations, although needing to have all the previous observations stored. The exponential average can offer similar performance, depending on the distribution of the samples, but it only needs to store the previous estimation. Therefore from an implementation point of view the exponential estimator is the one to choose.

In Figure 11 it is depicted the obtained RMSE while using both estimators for each of the channels. Two scenarios are considered, one where the channel is sensed each session, i.e. all samples are considered, and the case where the channel is sensed every second sensing session, i.e. half of the available samples are taken into account. $s_{reset}$ was set to 0.5, α was set to 0.01 and the $O_W$ was set to 40.

From Figure 11 it can be observed, as expected, that both estimators have similar performance, except in the Ch 2 during the interrupted sensing, where the RMSE was substantially higher for the exponential estimator. This was caused by the $s_{reset}$ value which was set as 0.5. Although it can be argued that by taking out the $s_{reset}$ the RMSE would be minimized, it needs to be considered since the estimation is used as channel bid for the scheduler and therefore needs to be updated in every sensing cycle. This will be further motivated in Section VI.

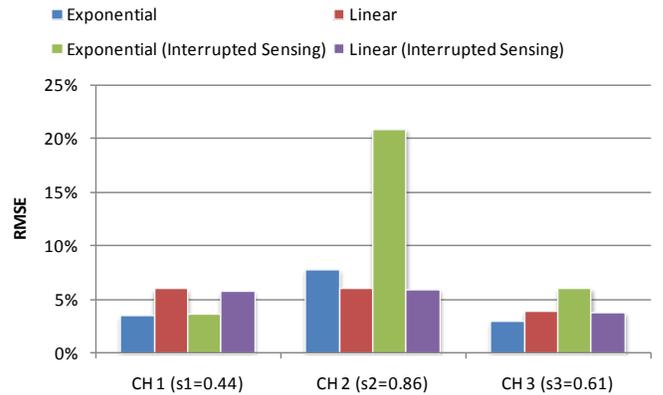

**Figure 11 - Estimators comparison through RMSE.**

## VI. SENSING ORCHESTRATION SCHEME

### A. Foundations of the Scheduling Scheme

In this section it is introduced the orchestrating scheme which distributes the sensing nodes across the spectrum along consecutive sensing sessions.

The proposed scheme is based on the Kelly scheme [16], here described according to [17], as follows. Consider the case where $M$ users are bidding for a share of a finite resource denoted as $C$. Here the scheme users are the channels being targeted for the sensing, while the resources to distribute are the sensing nodes. Each user *m* has associated a utility function, $A_m(d_m)$, which determines the monetary value of any resource allocation, $d_m$, to user *m*.

Let us consider the triple *(C, M, A)*, where *C > 0, M > 1*, and $\mathbf{A} = [A_1,...,A_M]$, as the utility system. The utility is measured in monetary units; therefore, if the user *m* receives a resource allocation $d_m$, it must pay a price $w_m$, thus receiving a net payoff $N_{PO}$, given by,

$$N_{PO} = A_m(d_m) - w_m \qquad (11)$$

Given any vector of utility functions **A** the maximization problem can be expressed as the maximization of the aggregate utility, and can be defined as the triple *(C, M, A)* such that:

$$\begin{aligned} maximize & \sum_{m=1}^{M} A_m(d_m) \\ subject\ to & \sum_{m=1}^{M} d_m \leq C \\ & d_m \geq 0 \end{aligned} \qquad (12)$$

where $d_m$ are the components of the non-negative vector of resource allocation **d**. This vector **d** can be computed, by considering that each user *m* submits a bid, denoted as $w_m$, to the resource manager, i.e. the cluster head. Then given the vector *w*, defined as $\mathbf{w} = [w_1,...,w_M]$, the resource manager chooses and allocates **d**, according to,

$$d_m(w_m) = \begin{cases} \frac{w_m}{\sum_{k=1}^{M} w_k} C, & if\ w_m > 0 \\ 0, & if\ w_m = 0 \end{cases} \qquad (13)$$

Considering that the users choose the bid based on the maximization of the $N_{PO}$, then according to [17], the allocation of the resources is fully efficient, reaching the maximum possible aggregate utility.

The challenge presented in this paper is to apply this scheduler both in a centralized and decentralized approach. In the centralized approach the cluster head is the one responsible for the orchestration, but when we follow a decentralized approach, there is not an entity/scheduler which computes and then assigns the resources to the channels accordingly. Therefore, to apply this scheduler to the decentralized case the Eq. (13) needs to be obtained indirectly, i.e. each of the sensing nodes will select which channel to sense based on the bids, and in the end the resource distribution will follow (13), although with some deviation.

### B. Centralized Orchestration Scheme Implementation

In Algorithm 2 it is shown the algorithm used to implement the centralized sensing orchestration scheme. To initialize the system the nodes sense a random channel in the first sensing session and from there one starts to estimate the $s_m$.

```
%Start Sensing Orchestration Session
Receive Sensing Results from Cluster Nodes
For (every channel) %Find expected allocated resources
    If (channel m has been sensed)
        Compute s_inst according with fusion rule
    Compute ŝ_m and obtain w_m
    Compute d_m(w_m)
    Allocate resources to next sensing session according to d_m
End
%End Sensing Orchestration Session
```
**Algorithm 2 – Centralized Orchestration Scheme.**

### C. Decentralized Orchestration Scheme Implementation

In Algorithm 3 it is shown the algorithm which accomplishes the distribution of the sensing nodes across the monitored channels in the decentralized case. The proposed algorithm occurs within each node after the period when the nodes report the sensing results, given in Figure 1-(a), and before the nodes perform the channel sensing, depicted in Figure 1-(b). To initialize the system the nodes sense a random channel in the first sensing session. By running this algorithm on all of the cluster nodes it is possible to obtain an approximation of Eq. (11).

```
%Start Sensing Orchestration Session
Receive Sensing Results from Cluster Nodes
For (every channel) %Compute the channel bids
    If (channel m has been sensed)
        Compute s_{m.inst} according to fusion rule
    Compute w_m, i.e. obtain ŝ_m
End
Normalize the channel bids, i.e. ||w_m|| = w_m / ∑_m^M w_m
Compute CDF, i.e.||w_m||_CDF = ||w_m|| + ||w_{m-1}||_CDF
Generate random variable, r ~ Uniform(0,1)
Select the minimal m for which true that r < ||w_m||_CDF
Select channel to sense with index m
%End Sensing Orchestration Session
```
**Algorithm 3 - Decentralized Orchestration Scheme.**

### D. Implementation Issues Comparison

Both of the proposed algorithms are implementable, although with some constraints, especially in the case of the decentralized algorithm. Indeed in the decentralized algorithm since it is not possible to know for certain what will be the number of sensing nodes which will be sensing a channel then it is not possible to apply the adaptive counting rule scheme for the data fusion. Therefore we only apply it in the centralized algorithm.

The centralized scheme will of course achieve a better performance than the decentralized, as expected. But the advantage of using the decentralized scheme is to give a higher robustness to the cooperative spectrum sensing scheme, since if in the centralized the scheme the central node stops working then the spectrum sensing also collapses, while in the decentralized in the case that some of the nodes withdraws from the cluster it will still be possible to continue the sensing.

The centralized and decentralized schemes complement each other, i.e. a possible way to increase the robustness of the cooperative spectrum sensing is to combine both schemes. So when the central node, by any reason, stops working the cluster can start working following the decentralized scheme, until another cluster head is elected. Through this is possible to achieve uninterrupted cooperative spectrum sensing.

## VII. Performance Evaluation

Here we evaluate the performance of the proposed orchestration algorithms, while comparing them with the Round Robin scheme performance, using of course the same data fusion scheme and channel estimator. In Table 1 are depicted the relevant simulation values.

Table 1 - Relevant simulation parameters.

| Parameter | Value |
|---|---|
| Estimation Method | Exponential |
| Mean Correlation Index | 0.5 |
| $\alpha$ | 0.01 |
| $s_{reset}$ | 0.5 |
| M | 20 |
| Channel Occupancy Model | Poisson On-Off Model |
| $P_{fa}$ | 0.05 |
| Sensing Sessions | 10000 |
| Local Detector Model | Energy detector |

Note that all the results presented, except for the theoretical ones, where obtained through the simulator developed for this study which implements the system design defined in Section III.

In Figure 12 and Figure 13 are depicted the $RMSE_{ME}(5)$ achieved by each of the orchestration schemes according to the number of local detectors and underlying data fusion rule. As expected the centralized scheme achieves a lower $RMSE_{ME}(5)$ since it allows for a more controlled distribution of the sensing users. Also in the decentralized scheme evaluation it was not considered the adaptive counting fusion rule due to a lack of not being able to control exactly the number of nodes to sense a channel, as explained in the sub-section VI.D.

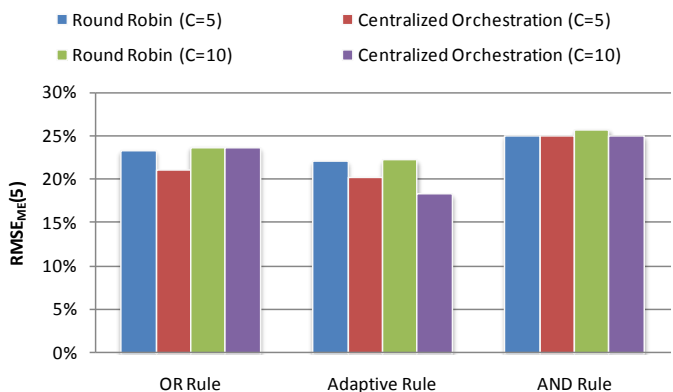

Figure 12 - Centralized scheme $RMSE_{ME}(5)$ evaluation.

## VIII. Conclusion

In this paper it we proposed a robust cluster based cooperative spectrum sensing scheme to be used in an Ad-hoc cluster network in a disaster relief context. The proposed scheme allows the distribution of the sensing nodes to be adapted according to the channel mutating conditions, therefore maximizing the correct identification of spectrum holes.

The proposed scheme system design was fully described and its underlying steps explained. We explained the signaling flow to achieve the proposed scheme. We gave special focus on three steps of the proposed scheme, i.e. data fusion, channel state estimation, and sensing orchestration.

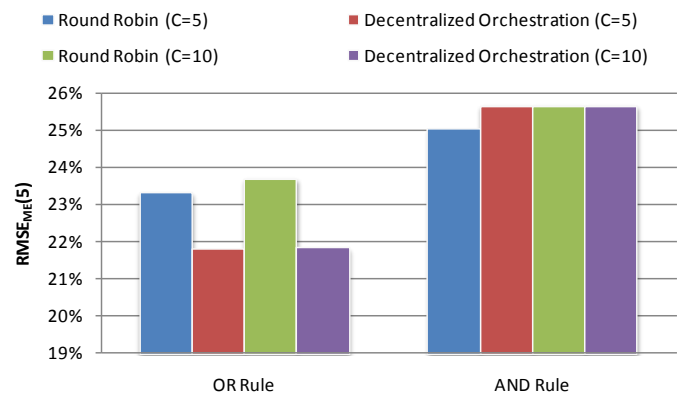

Figure 13 - Decentralized scheme $RMSE_{ME}(5)$ evaluation.

As per data fusion method, we proposed an adaptive counting rule. This allows adapting the decision threshold dynamically according to the correlation experienced by the underlying sensors, while not having implicit information about the correlation, i.e. acting therefore as an adaptive feedback mechanism.

As per channel state estimation we evaluated two schemes, the Exponential Moving Average and the Linear Moving Average estimator. It was concluded that the exponential estimator allows for a simpler implementation while achieving almost the same performance as the linear moving average estimator, which is more expensive to implement.

As per the spectrum sensing orchestration scheme, we proposed a centralized and decentralized orchestration schemes. From the performance evaluation it was concluded that the centralized scheme achieves a lower $RMSE_{ME}(5)$ than the decentralized. From a system design perspective the decentralized scheme is more robust since in the case one of that any of the cluster nodes leaves the cluster the decentralized scheme continues to work, while in the centralized scheme, if the cluster head withdraws from the cluster the cooperative spectrum sensing collapses immediately. So it was proposed in the future to combine both of the schemes which will potentially allow for a more robust cooperative sensing scheme.

The future steps of this work will be to extend the proposed orchestrating mechanism to a multi cluster scenario, where it will be studied how to support mobility between clusters as well continuation of the knowledge base continuity of the estimated statistics in the network. Also it should be studied the effect on the performance of the proposed schemes when the Control Channel is imperfect.

## Acknowledgment

The authors would like to thank Professor Kwang-Cheng Chen, from the Department of Electrical Engineering National Taiwan University, for the fruitful discussion about his more recent research results on cooperative spectrum sensing.